\documentclass[11pt]{article}

\usepackage[preprint]{acl}

\usepackage{times}
\usepackage{latexsym}
\usepackage[T1]{fontenc}
\usepackage[utf8]{inputenc}
\usepackage{microtype}
\usepackage{inconsolata}

\usepackage{graphicx}
\usepackage{amsmath}
\usepackage{amssymb}
\usepackage{amsfonts}
\usepackage{amsthm}
\usepackage{booktabs}
\usepackage{multirow}
\usepackage{threeparttable}
\usepackage{colortbl}
\usepackage{array}
\usepackage{makecell}
\usepackage{tabulary}
\usepackage{arydshln}
\usepackage{wrapfig}
\usepackage{epsfig}
\usepackage{subfig}
\usepackage{dblfloatfix}
\usepackage[export]{adjustbox}

\usepackage[export]{adjustbox}
\usepackage{pgfplots}
\pgfplotsset{compat=1.18}

\usepackage{tikz}
\usepackage{pgfplots}
\pgfplotsset{compat=1.18}

\usepackage{enumitem}
\usepackage{siunitx}
\usepackage{bm}
\usepackage{nicefrac}
\usepackage{pifont}
\usepackage{xspace}
\usepackage{soul}

\usepackage{algorithm}
\usepackage{algorithmic}
\usepackage{listings}

\usepackage{xcolor}
\usepackage{url}
\usepackage{hyperref}
\usepackage[capitalize]{cleveref}
\usepackage{lipsum}
\usepackage{enumitem}

\newcommand{\ie}{\textit{i}.\textit{e}.}
\newcommand{\eg}{\textit{e}.\textit{g}.}

\newcommand{\oursname}{\texttt{PspMAS}}
\newcommand{\ours}{\oursname\xspace}

\definecolor{delay}{RGB}{230,94,42}
\definecolor{mywarning}{RGB}{233,144,61}
\definecolor{mygray}{gray}{.9}
\definecolor{ggray}{RGB}{127,127,127}
\definecolor{reda}{RGB}{192,0,0}
\definecolor{redb}{RGB}{217,148,143}
\definecolor{myyellow}{RGB}{190,144,0}
\definecolor{mygreen}{RGB}{80,100,40}
\definecolor{myblue}{RGB}{30,90,100}
\definecolor{mypurple}{RGB}{135,82,155}
\definecolor{oriHOI}{RGB}{236,90,67}
\definecolor{HOI}{RGB}{144,67,57}
\definecolor{obj}{RGB}{247,206,84}
\definecolor{bestred}{RGB}{192,0,0}

\makeatletter
\newcommand{\thickhline}{
  \noalign{\ifnum0=`}\fi \hrule height 1pt
  \futurelet \reserved@a \@xhline
}
\makeatother

\newcolumntype{y}[1]{>{\raggedright\arraybackslash}p{#1pt}}
\newcolumntype{z}[1]{>{\raggedleft\arraybackslash}p{#1pt}}

\theoremstyle{definition}

\crefname{section}{Sec.}{Secs.}
\Crefname{section}{Section}{Sections}
\crefname{subsection}{Sec.}{Secs.}
\Crefname{subsection}{Section}{Sections}
\crefname{figure}{Fig.}{Figs.}
\Crefname{figure}{Figure}{Figures}
\crefname{table}{Tab.}{Tabs.}
\Crefname{table}{Table}{Tables}
\crefname{equation}{Eq.}{Eqs.}
\Crefname{equation}{Equation}{Equations}

\title{Scaling Multi-Agent Systems with Prospect-State Propagation}

\author{
  Zhimei Chen\textsuperscript{1}\thanks{Equal contribution.} \quad
  Mu Chen\textsuperscript{2}\footnotemark[1]\thanks{Corresponding author.} \quad
  Fakhri Karray\textsuperscript{2,3} \\
  \textsuperscript{1}Independent Researcher \\
  \textsuperscript{2}Mohamed bin Zayed University of Artificial Intelligence \\
  \textsuperscript{3}University of Waterloo \\
  \texttt{mu.chen@mbzuai.ac.ae} \\
}

\begin{document}
\maketitle

\begin{abstract}
Current LLM-based multi-agent systems (MAS) periodically compress intermediate states to reduce inference-time token consumption, thereby attempting to incorporate more agents. However, naive scaling strategies face challenges. For example, in economic simulations, large-scale MAS typically discard semantically rich economic states, \ie, agent behavioral trajectories, which are key drivers of macroeconomic fluctuations. In this paper, we reveal a phenomenon in which agent heterogeneity gradually decreases during simulation, and propose \textbf{P}rospect-\textbf{S}tate \textbf{P}ropagation for \textbf{M}ulti-\textbf{A}gent \textbf{S}ystems (\ours). Inspired by prospect theory, \ours decouples each agent's micro state into a compact Prospect State and an expressive Semantic State. The former records psychological traces through a lightweight, parallelizable propagator and continuously injects heterogeneity into the system. The latter leverages the strong perception, reasoning, planning, and decision-making abilities of LLMs. These two components work complementarily, providing a scalable LLM-based multi-agent simulation solution.
\end{abstract}

\section{Introduction}
\label{sec:intro}
\begin{quote}
\it \small
Humans evaluate potential gains and losses relative to a specific reference point rather than focusing on absolute wealth, ..., is Prospect Theory.
 \\
\mbox{}\hfill -- Daniel and Amos (1979)
\vspace{-3pt}
\end{quote}

From Adam Smith's \textit{“invisible hand”} in the 18th century to the \textit{behavioral economics} revolution of the 20th century, the study of Macroeconomics has increasingly highlighted the importance of modeling the complexity of human behavior. Traditional representative-agent models \cite{kirman1992, blanchard2017, christiano2005} largely circumvent this difficulty by assuming a “perfect world” populated by a single average consumer or firm. Such assumptions, however, eliminate distributional heterogeneity and the non-linear aggregate fluctuations \cite{gabaix2011granular} it generates, limiting the ability to replicate typical macroeconomic events (\eg, crises and recessions).

   \begin{figure}[t]
      \begin{center}
     \includegraphics[width=1.\linewidth]{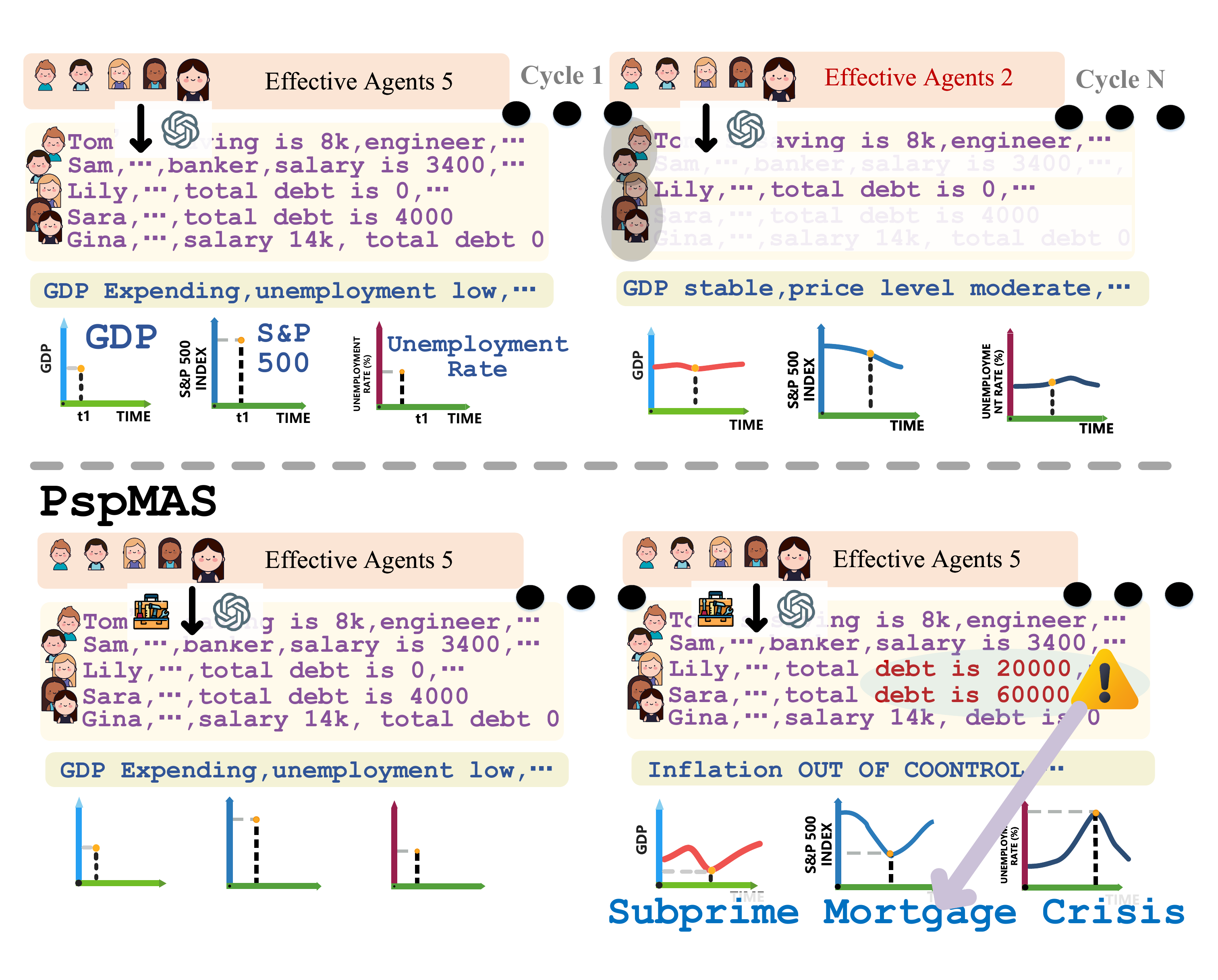}
          \end{center}
      \vspace{-8pt}
      \captionsetup{font=small}
\caption{\small
The top sub-figure shows a traditional multi-agent simulation system. After a long-horizon simulation, at cycle $N$, the heterogeneity of agents decreases, and the macroeconomic fluctuations are consequently weakened.
The bottom sub-figure shows \ours. By preserving heterogeneity, the sharp increase in personal mortgage debt for some agents dominates the description of their micro states, thereby facilitating the emergence of macroeconomic phenomena (e.g., the subprime mortgage crisis).
}
      \label{fig:1}
            \vspace{-0pt}

    \end{figure}

Agent-based modeling (ABM) offers a bottom-up solution \cite{farmer2009economy, tesfatsion2006handbook}. By simulating the behaviors and interactions of heterogeneous agents, ABM allows stylized facts (\eg, GDP, inflation) to emerge from the micro level. Rule-based ABMs maintain a large number of agents with their economic microstate (\eg, income, savings, debt, employment status, wage) that are updated via hand-crafted functions, one representative microstate in the form of key-value pair is \!\! \hspace{0.03em}
\raisebox{-0.12em}{
\includegraphics[height=0.9em]{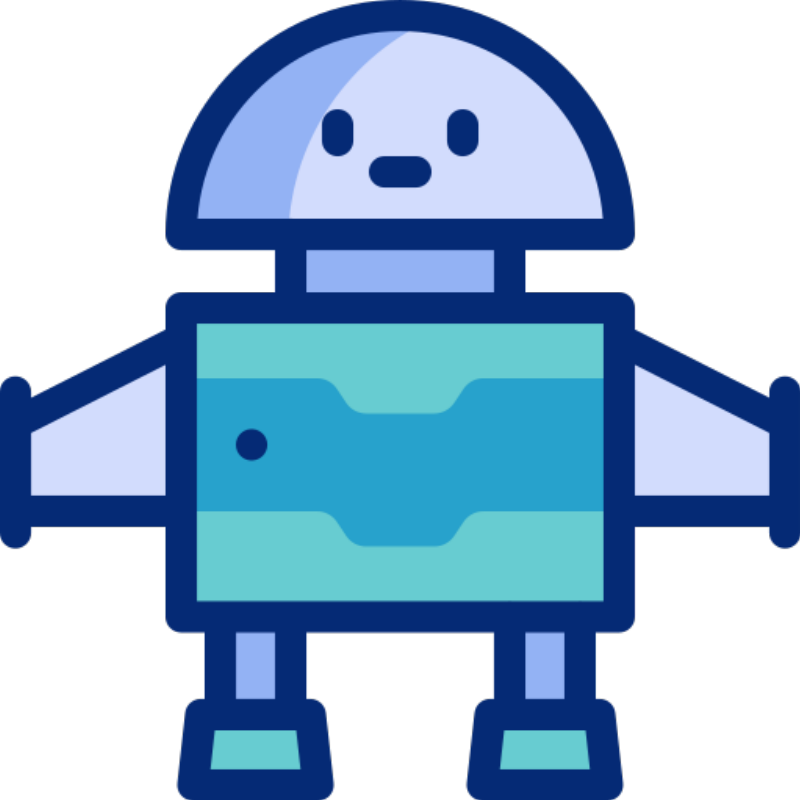}
}{\ttfamily\color{myblue}
`agent:1,\! income:3200,\! saving:9000,\! unemployment\! status:0,\! ...'}. Such micro states can be batched, parallelized, and vectorized, enabling hardware-efficient simulation. However, the pre-defined rules fail to capture real psychological processes, ultimately hindering the replication of key macroeconomic dynamics driven by complex human cognition.

Large language models (LLMs) have recently shown remarkable capabilities in natural language understanding  \cite{openai2023gpt4,liang2022holistic,zheng2023judging}, role-playing \cite{wang2024rolellm}, and contextual reasoning \cite{yao2023react,zhu2024can,dalal2024inference}, making them attractive for social simulation \cite{horton2023,argyle2023,li2024econagent,jia2024can,binz2025foundation,wang2024rolellm,zou2025survey,li2023camel}. The microstate of an LLM-based agent is no longer a fixed set of numbers but a rich semantic description, for example: \raisebox{-0.12em}{
\!\!\!\includegraphics[height=0.9em]{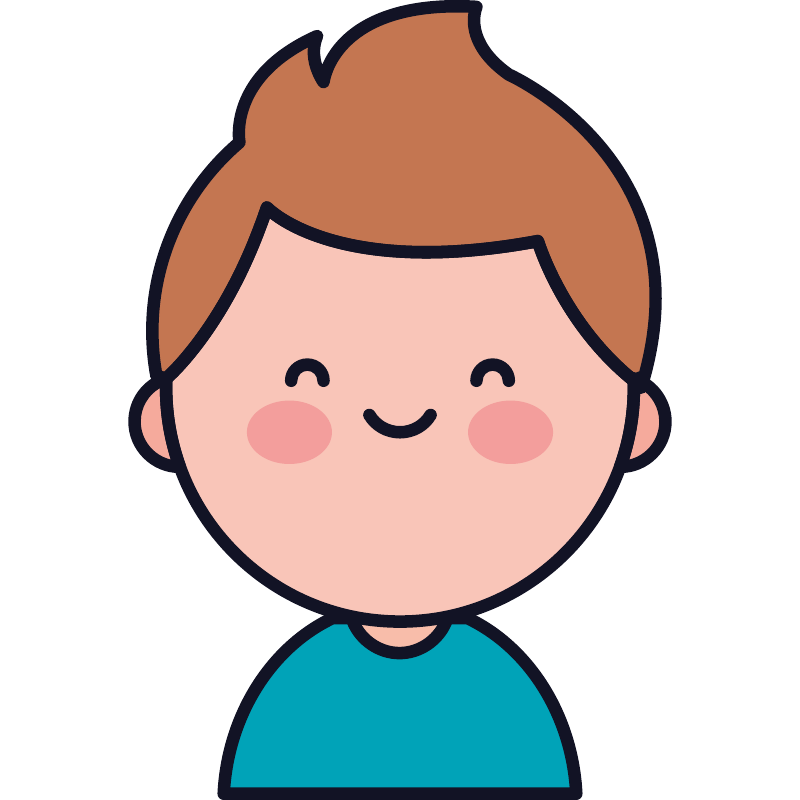}
}\!\!
{\ttfamily\color{mypurple}
`You're\!\! Tom,\!\! an\!\! engineer\!\! aged\!\! 25.\!\! Your\!\! saving\! is\!\! 8000,\! your\!\! salary\! is\! 3200,\! and\! total\! debt\! is\! 0,\! ...,\! CPI\! indicates\! current\! economy\! is in a high-inflation cycle.\! What\! level\! of\! expenditure\! do\! you\! plan\! for next period?'} Such semantic-rich microstates enable sufficient behavioural heterogeneity, which is a key driver of macroeconomic fluctuations. But long descriptive profiles of agents during long-horizon simulation consume substantial inference-time token budgets, hindering scalability to large agent populations. Existing multi-agent systems resort to periodic reflection~\cite{park2023generative,shinn2023reflexion}, compressing economic histories into high-level summaries. While effective in multi-turn conversations\cite{qian2024chatdev,yang2024swe,wu2024autogen,wang2023describe} or code generation tasks, such semantic compression strategies discard the temporal order and intermediate states that are essential for economic decision-making. From the perspective of behavioral economics, human behavior is highly sensitive to early economic fluctuations, and even small changes can exert lasting influences on subsequent choices. Repeated summarization across simulation cycles tends to decrease the dynamic heterogeneity of agents, leading to behavioral homogenization---a phenomenon called \emph{action convergence}. As illustrated in the top part of Fig. \ref{fig:1}, a traditional system injects rich descriptions into all agents at system initialization, thereby creating a large nominal population of heterogeneous agents. However, as the simulation proceeds, the micro states of agents are compressed into high-level summaries on a recurring basis. This process gradually removes individual behavioral differences and greatly reduces diversity among agents. Consequently, macroeconomic fluctuations that should emerge from heterogeneous local distress become attenuated. In macroeconomics, risk signals with genuine predictive value often first appear in the local states of a small subset of agents. If state descriptions are dominated by lagging macro variables such as the unemployment rate, these localized abnormalities can be diluted or erased during semantic compression. As a result, most agents are guided toward a homogenized common context in which the economy still appears broadly healthy, suppressing the micro-to-macro propagation through which crises emerge. For example, in the buildup to the subprime mortgage crisis, early risks were not immediately reflected in aggregate labor-market indicators. Instead, stress first accumulated in agents with average wages but high debt burdens, such as {\ttfamily\color{mypurple} agent\! Sara\!} 
\raisebox{-0.12em}{
\includegraphics[height=0.9em]{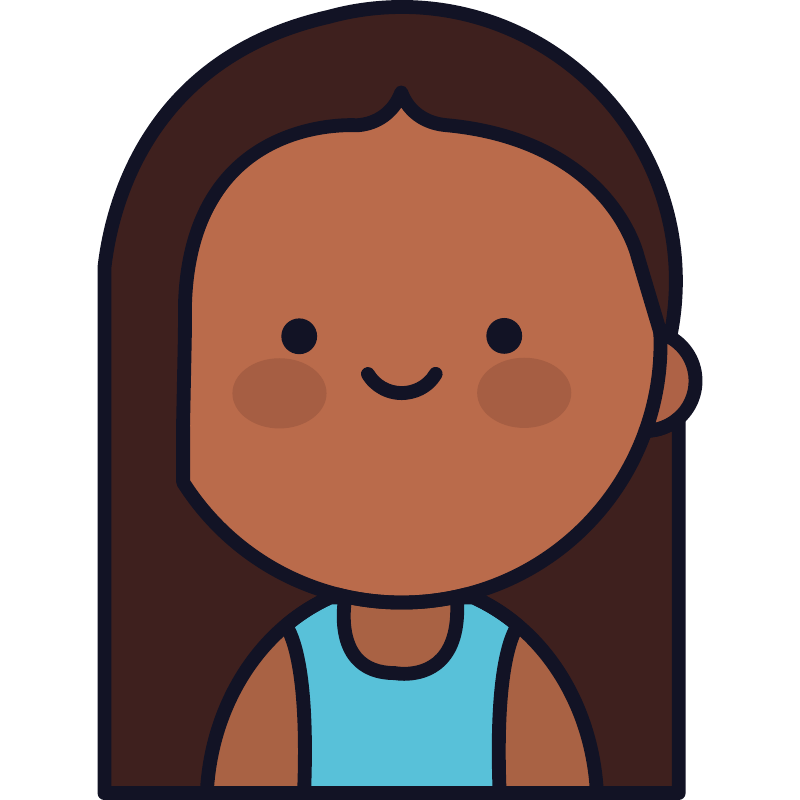}
}shown in the bottom part of Fig. \ref{fig:1}. Such agents can transmit early default risk and other key local signals that eventually develop into the subprime mortgage crisis. If these loan-related micro states are compressed away, the system loses the heterogeneous mechanisms through which localized distress propagates into broader macroeconomic fluctuations.

Based on the above observations, we propose 
\textbf{P}rospect-\textbf{S}tate \textbf{P}ropagation for 
\textbf{M}ulti-\textbf{A}gent \textbf{S}ystems (\ours). 
\ours decouples each agent's micro state into two complementary parts: a compact \emph{Prospect State} and an expressive \emph{Semantic State}. 
The design of \emph{Prospect State} is inspired by prospect theory, a classical theory in behavioral economics for characterizing human decision-making under gains, losses, and uncertainty. 
It maintains key psychological variables, including reference point, loss aversion, and probability weighting, and updates them through a fast propagator. \emph{Prospect State} provides the long-term micro-level heterogeneity necessary for economic emergence.
\emph{Semantic State} leverages the perception, reasoning, planning, and decision-making abilities of LLMs to generate human-like actions. 
By combining these two states, \ours avoids relying on long textual state descriptions, making the system easier to scale, while also mitigating the heterogeneity loss caused by long-horizon semantic compression. 
Our contributions are summarized as follows:
\begin{itemize}[leftmargin=*, noitemsep]
    \item We reveal an inherent heterogeneity-loss problem in MAS for economic simulation, where periodical semantic compression gradually reduces the effective number of behaviorally distinct agents.
    \item We incorporate prospect theory, preserving long-term heterogeneity.
    \item We introduce a scalable LLM-based multi-agent simulation solution.
\end{itemize}

\section{Method}
\label{sec:method}This section introduces \ours, a framework for preserving agent heterogeneity while scaling LLM-based multi-agent systems over long-horizon simulation. The key idea is to decouple each agent's micro state into two complementary parts: a Markovian Prospect State updated by a lightweight Prospect-State Propagator, and a Semantic State updated periodically by an LLM. We first provide background on multi-agent systems for macroeconomic simulation (Section~\ref{subsec:preliminary}), then introduce the simulation loop, including economic state construction, prospect-state propagation, economic state read, and system scaling. Implementation details are given in Section~\ref{subsec:implementation}.

\subsection{Preliminary}
\label{subsec:preliminary}
In a Multi-agent Systems (MAS) \cite{li2024econagent, feng2025simcity}, agents perceive the environment and take actions to achieve goals; their behavior is not hard-coded but emerges from situated interactions with other agents and the environment. We consider a MAS for macroeconomic simulation populated by $N$ heterogeneous agents, indexed by $\mathcal{I}=\{1,2,\ldots,N\}$. The corresponding agent set is $\{A_i\}_{i\in\mathcal{I}}$. Each agent is equipped with LLM-powered perception, reasoning, planning, and decision-making abilities. During simulation, agents interact, collectively constructing a macroeconomy, which in turn updates their perceptions and influences subsequent actions.

\subsection{\ours}
\label{subsec:ours}

\paragraph{Economic State Construction.}
\label{subsec:economic_state}
Micro differences of agents are a key driver of macroeconomic fluctuations. The combination of agents' micro states and the environment's macro state constitutes the Economic State of the simulation.  

\textit{Micro State}. Each agent $A_i$ at time $t$ maintains an independent micro state capturing its internal behavioral state and economic history. In traditional rule-based ABMs, the micro state is a fixed-dimensional structured vector \cite{kuroki2025reimagining}. In LLM-based simulations, the micro state typically comprises a compact numerical component and a rich semantic component \cite{li2024econagent}. In \ours, we write the micro state as
\begin{equation}
    \mathcal{M}_i^t = \bigl( s_i^t,\; \sigma_i^t,\; e_i^t \bigr), \label{eq:microstate}
\end{equation}
where $s_i^t\in\mathcal{P}$ is the Prospect State, $\sigma_i^t\in\mathcal{M}_{\text{sem}}$ is the Semantic State, and $e_i^t$ denotes economic bookkeeping variables (e.g., wealth, debt, tax paid, wage, realized consumption, employment status) that are updated by the environment via market clearing. In Particular, the two central branches of \ours are $s_i^t$ and $\sigma_i^t$: the former preserves compact psychological traces, while the latter supports expressive natural-language reasoning.

\textit{Macro State}. The overall condition of the economic environment at time $t$ is described by a macro state $\mathcal{S}^t$, typically including price level $P^t$, interest rate $r^t$, unemployment rate $u_{\mathrm{emp}}^t$, and total GDP $Y^t$. Formally, $\mathcal{S}^t \in \mathcal{Y}$. The macro state reflects the aggregation of all agents' actions through market mechanisms, forming a closed feedback loop \cite{li2024econagent, yang2026twinmarket}.

\textit{Agent Action}. At each time step $t$, each agent $A_i$ selects an action $a_i^t$ from its action space $\mathbb{A}_i$ based on its current micro state $\mathcal{M}_i^t$ and the macro state $\mathcal{S}^t$. Following macroeconomic simulation literature, the action space includes two fundamental decisions: whether to work ($l_i^t \in \{0,1\}$) and what fraction of available funds to spend on consumption ($p_i^t \in [0,1]$) \cite{li2024econagent, feng2025simcity}. The action space can be expressed as
\begin{equation}
    a_i^t = (l_i^t, p_i^t) \in \mathbb{A}_i = \{0,1\} \times [0,1].
\end{equation}

    \begin{figure*}[t]
        \begin{center}
            \includegraphics[width=1\linewidth]{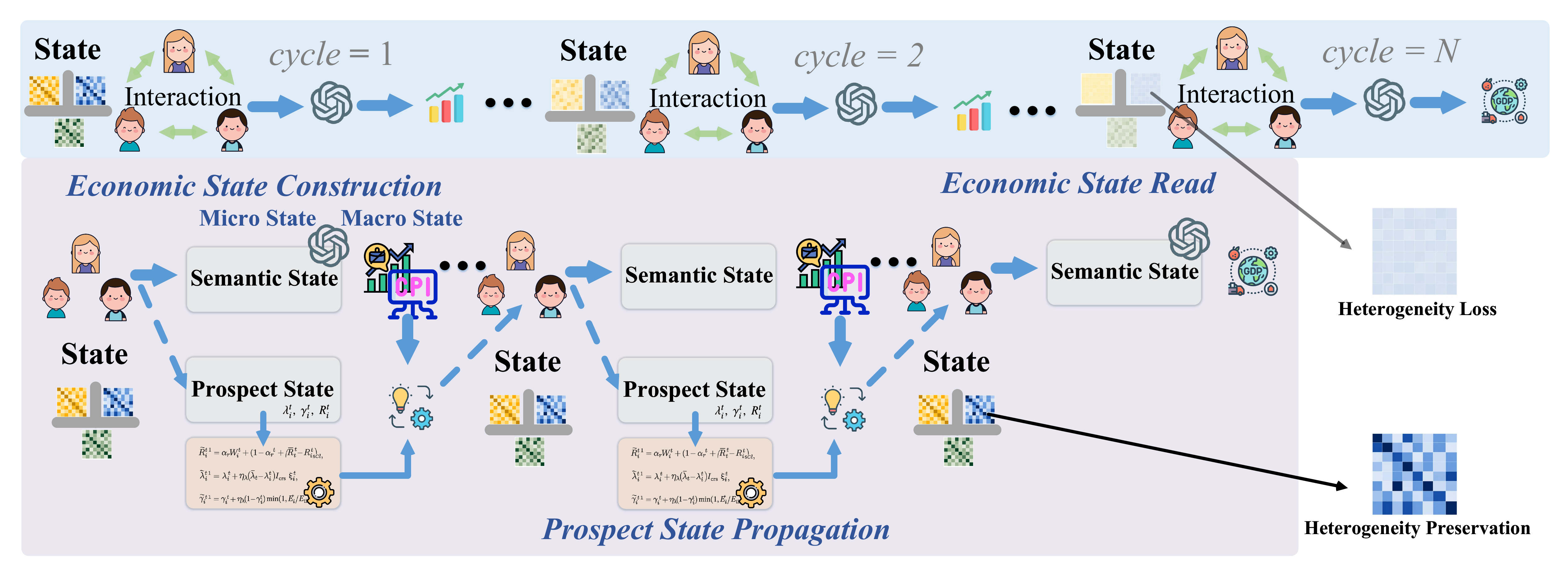}
            \end{center}
        \captionsetup{font=small}
        \caption{\small{Overview of the major components of \ours. \ours decouples each agent's micro state into two complementary branches: a compact Prospect State and an expressive Semantic State. The Prospect State is continuously updated by a lightweight Prospect-State Propagator, preserving psychological traces and injecting long-term heterogeneity into the agent population. The Semantic State is periodically refreshed by the LLM to support perception, reasoning, planning, and human-like decision-making. \ours preserves heterogeneous micro dynamics while enabling scalable long-horizon macroeconomic simulation.}
        }
        \label{fig:2} 
        
    \end{figure*}

\paragraph{Prospect-State Propagation.}
\label{subsec:prospect_state_propagation}
We incorporate prospect theory \cite{kahneman2013prospect} into the micro state to capture bounded rationality in decision-making. The theory highlights three key features:
- \textit{Reference dependence}: decisions are made based on gains and losses relative to a psychological reference point (e.g., initial wealth), rather than absolute wealth.
- \textit{Loss aversion}: losses are felt approximately twice as strongly as equivalent gains (i.e., $\lambda \approx 2.25$).
- \textit{Probability weighting}: people tend to overweight small probabilities and underweight moderate to high probabilities, leading to distorted subjective probabilities.

Based on these insights, we define the Prospect State of agent $i$ at time $t$ as a Prospect-State Vector:
\begin{equation}
    s_i^t = \bigl( \lambda_i^t,\; \gamma_i^t,\; R_i^t \bigr) \in \mathcal{P} \subset \mathbb{R}^3. \label{eq:prospect_state}
\end{equation}
Here $R_i^t$ is the dynamic reference point, $\lambda_i^t$ the loss-aversion coefficient, and $\gamma_i^t$ the probability-weighting parameter. Other economic quantities, such as wealth, income, realized consumption, and unmet demand, are stored as accounting variables or immediate outcomes rather than as dimensions of the Prospect-State Vector.

Let $o_i^t$ denote the immediate individual outcome produced by market clearing at time $t$ (e.g., realized labor income, realized consumption, savings change, and unmet demand), and let $\mathcal{S}^{t+1}$ denote the realized macro state after clearing. The Prospect State update is Markovian:
\begin{equation}
    \tilde{s}_i^{t+1} = \Phi(s_i^t, o_i^t, \mathcal{S}^{t+1}), \qquad
    s_i^{t+1} = \Pi_{\mathcal{P}}(\tilde{s}_i^{t+1}),
    \label{eq:prospect_state_update}
\end{equation}
where $\Pi_{\mathcal{P}}$ projects the updated Prospect-State Vector onto a predefined valid domain $\mathcal{P}$. This projection step, rather than the coefficient ranges alone, keeps the Prospect State bounded throughout long-horizon simulation.

The Markovian Prospect-State Propagator $\Phi$ consists of deterministic, parallelizable lightweight rules:
\begin{equation}
\begin{aligned}
\tilde{R}_i^{t+1} &= \alpha_r W_i^t + (1-\alpha_r) R_i^t + \beta_r (\bar{R}_t^c - R_i^t) I_{\mathrm{scl}}^t, \\
\tilde{\lambda}_i^{t+1} &= \lambda_i^t + \eta_\lambda (\bar{\lambda}_t - \lambda_i^t) I_{\mathrm{crs}}^t + \xi_i^t, \\
\tilde{\gamma}_i^{t+1}\! &=\! \gamma_i^t + \eta_\gamma (1-\gamma_i^t) \min(1, E_i^t/E_{\mathrm{thr}}) \!-\! \zeta I_{\mathrm{shk}}^t,
\end{aligned}
\label{eq:psp_rules}
\end{equation}
where $W_i^t$ is total wealth, $\bar{R}_t^c$ the cohort-average reference point, and $I_{\mathrm{scl}}^t$, $I_{\mathrm{crs}}^t$, and $I_{\mathrm{shk}}^t$ are indicator flags for social comparison, crisis, and shock, respectively. The variable $E_i^t$ denotes a recent experience intensity derived from the immediate outcome $o_i^t$, and $E_{\mathrm{thr}}$ is a normalization threshold. After computing Eq.~\ref{eq:psp_rules},  the projection in Eq.~\ref{eq:prospect_state_update} is then applied component-wise to obtain the next Prospect-State Vector, which is denoted by $s_i^{t+1}$.

The Semantic State $\sigma_i^t$ is a textual summary of recent events and agent reflections, updated \emph{periodically} (every $K$ steps) by an LLM. The LLM receives the current Prospect State $s_i^t$, the current accounting variables $e_i^t$, a short history of the most recent $K$ raw events, and the previous Semantic State $\sigma_i^{t-K}$, then produces a new $\sigma_i^t$. This branch provides perception, reasoning, and human-like decision-making. The two branches operate in parallel: the Prospect State evolves continuously at low cost, preserving long-term path dependence; the Semantic State is refreshed on a longer timescale, injecting rich behavioral patterns without prohibitive token costs.

\paragraph{Economic State Read.}
\label{subsec:economic_read}
After all agents have taken their actions, the environment updates the Economic State via a \textit{market clearing function} $\Psi$:
\begin{equation}
(\mathbf{M}^{t+1}, \mathcal{S}^{t+1}, \mathbf{o}^t) = \Psi\bigl( \mathbf{a}^t, \mathbf{M}^t, \mathcal{S}^t \bigr), \label{eq:transition}
\end{equation}
where $\mathbf{M}^t = (\mathcal{M}_i^t)_{i\in\mathcal{I}}$, $\mathbf{a}^t = (a_i^t)_{i\in\mathcal{I}}$, and $\mathbf{o}^t=(o_i^t)_{i\in\mathcal{I}}$ contains the individual realized outcomes used by the Prospect-State Propagator. In our implementation, $\Psi$ aggregates intended consumption and labor supply. If total intended demand exceeds total production, goods are rationed proportionally; individual unmet demand $q_i^t$ is recorded as part of $o_i^t$ and can affect the next Prospect State update through the experience intensity $E_i^t$. Prices and wages adjust according to the imbalance (e.g., demand $>$ supply $\rightarrow$ price increase). The government collects progressive taxes and redistributes all revenue equally. Annually, the central bank sets the interest rate using a Taylor rule, and savings earn that interest. This transition closes the loop, feeding macro conditions back into agents' future states, thus cycling repeatedly.

\paragraph{System Scaling.}
\label{subsec:scaling}
\ours is designed to scale efficiently to large agent populations while preserving behavioral diversity.

\textit{Computational Scalability}. The Prospect State is updated through lightweight and highly parallelizable propagation rules, avoiding expensive long-context reasoning during most simulation steps. Meanwhile, the Semantic State is refreshed only periodically rather than continuously, substantially reducing the number of costly LLM inference calls while keeping prompts compact over long simulation horizons. Since both state-update branches operate independently across agents, the overall system scales efficiently with the number of agents and simulation steps, enabling large-scale macroeconomic simulations with thousands of agents under practical computational budgets.

\textit{Heterogeneity Quantification}. To measure whether diversity is maintained at scale, we construct three population-level matrices: a Trajectory Matrix, a Prospect Matrix, and a fused Heterogeneity Matrix. The Trajectory Matrix captures realized behavioral diversity from recent action sequences, while the Prospect Matrix captures latent path-dependent psychological diversity from Prospect-State Vectors. The fused matrix combines both views and is used to compute the effective number of distinct behavioral modes.

\raisebox{-0.12em}{
\includegraphics[height=0.9em]{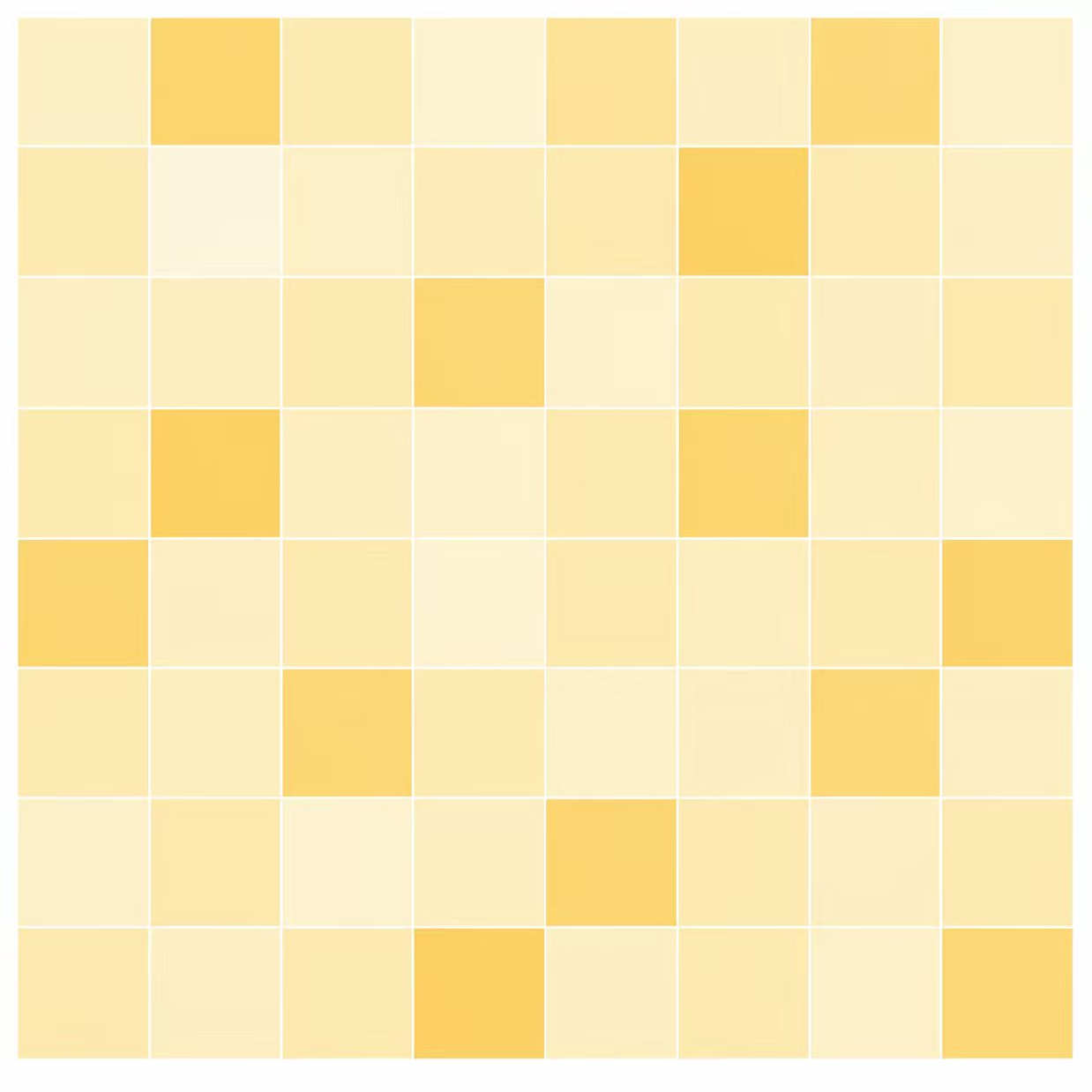}
}\textit{Trajectory Matrix.} Let
\begin{equation}
\mathbf{z}_i^t = (p_i^{t-L+1}, l_i^{t-L+1}, \ldots, p_i^{t}, l_i^{t})
\end{equation}
be the action sequence of agent $i$ over the last $L$ steps ($L=12$). After normalizing $\mathbf{z}_i^t$ over agents, we construct the Trajectory Matrix as an RBF similarity kernel:
\begin{equation}
H_{\mathrm{traj},ij}^t
=
\exp\!\left(
-\frac{
\|\hat{\mathbf{z}}_i^t-\hat{\mathbf{z}}_j^t\|_2^2
}{
2\tau_{\mathrm{traj}}^2
}
\right),
\label{eq:traj_matrix}
\end{equation}
where $\hat{\mathbf{z}}_i^t$ is the normalized trajectory feature and $\tau_{\mathrm{traj}}$ is a bandwidth parameter. This matrix is large when two agents have similar recent labor and consumption trajectories.

\raisebox{-0.12em}{
\includegraphics[height=0.9em]{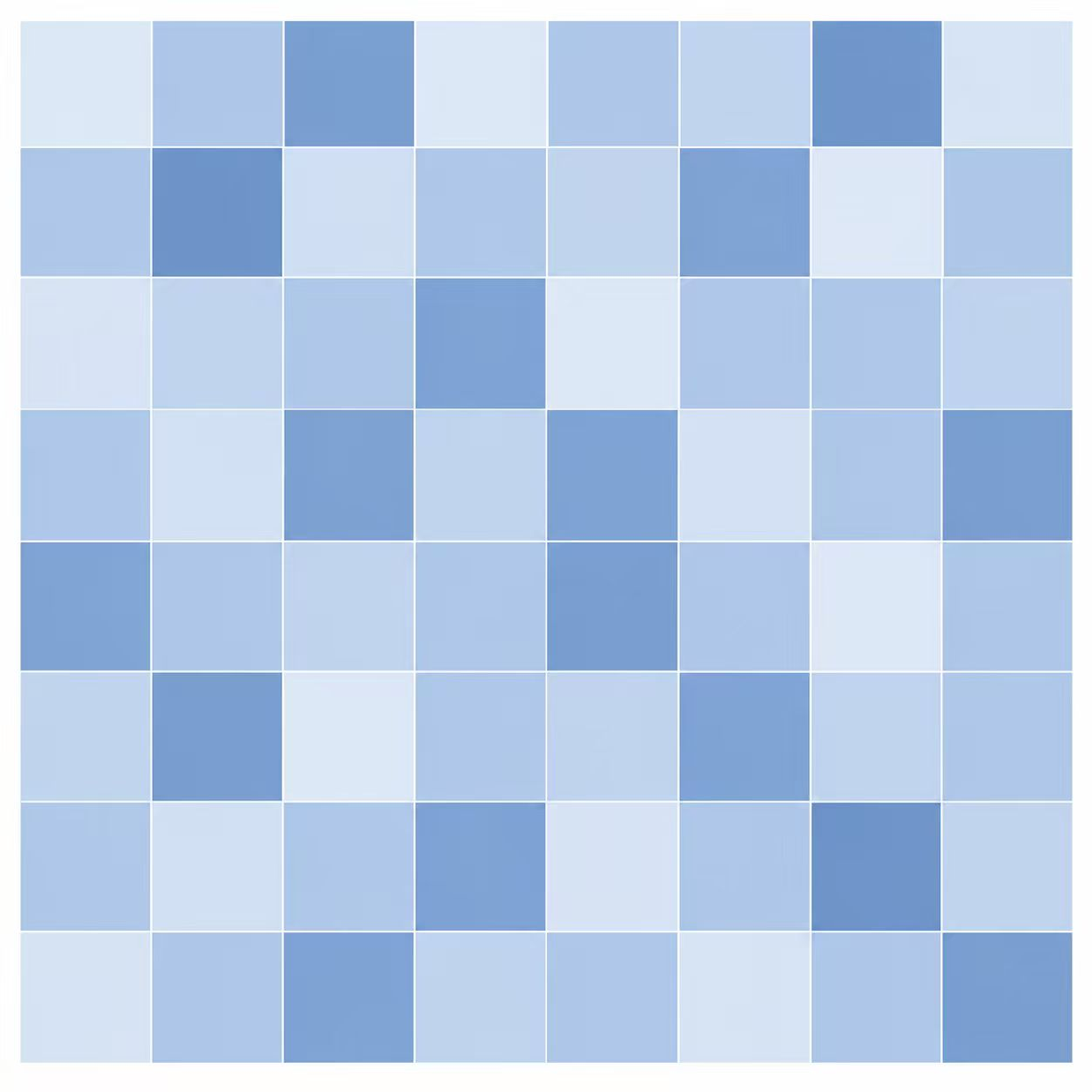}
}\textit{Prospect Matrix.} The Prospect Matrix is constructed from the Prospect-State Vector
\begin{equation}
s_i^t = (\lambda_i^t, \gamma_i^t, R_i^t).
\end{equation}
After normalizing $s_i^t$ over agents, we define
\begin{equation}
H_{\mathrm{pros},ij}^t
=
\exp\!\left(
-\frac{
\|\hat{s}_i^t-\hat{s}_j^t\|_2^2
}{
2\tau_{\mathrm{pros}}^2
}
\right),
\label{eq:pros_matrix}
\end{equation}
where $\hat{s}_i^t$ is the normalized Prospect-State feature and $\tau_{\mathrm{pros}}$ is a bandwidth parameter. This matrix captures whether two agents have similar path-dependent psychological states, even when their recent actions appear similar.

\raisebox{-0.12em}{
\includegraphics[height=0.9em]{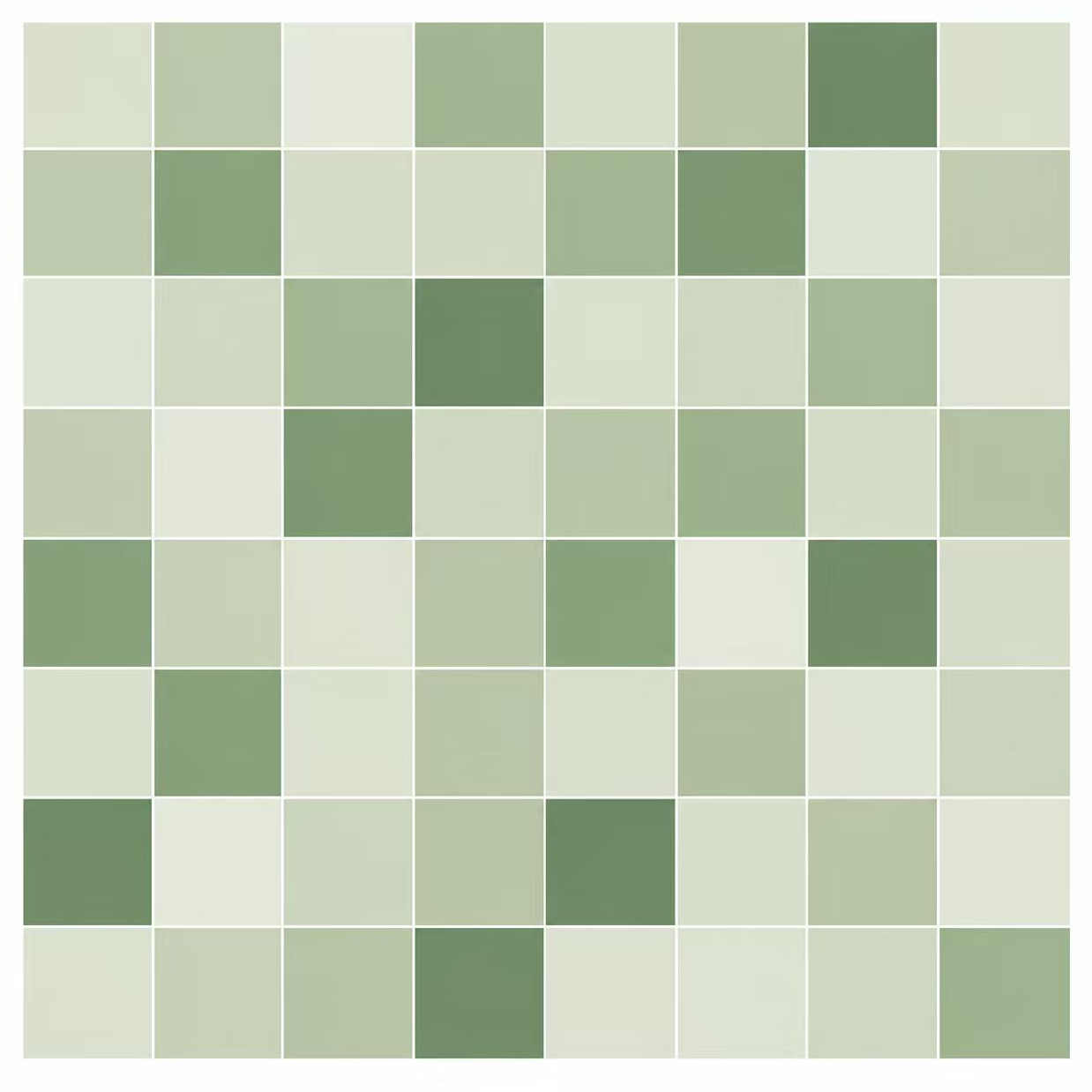}
}\textit{Fused Heterogeneity Matrix.} We combine the two matrices with a balance coefficient $\alpha\in[0,1]$:
\begin{equation}
H_{\mathrm{fuse}}^t
=
\alpha H_{\mathrm{traj}}^t
+
(1-\alpha)H_{\mathrm{pros}}^t.
\label{eq:fused_matrix}
\end{equation}
Because both $H_{\mathrm{traj}}^t$ and $H_{\mathrm{pros}}^t$ are RBF kernels, the fused matrix is positive semidefinite. Therefore, its spectrum can be used as a stable diversity measure.

Let $\mu_1^t,\mu_2^t,\ldots,\mu_N^t$ be the eigenvalues of $H_{\mathrm{fuse}}^t$, and normalize them as
\begin{equation}
\tilde{\mu}_k^t
=
\frac{\mu_k^t}{\sum_{j=1}^N \mu_j^t}.
\end{equation}
We define the spectral entropy as
\begin{equation}
\mathcal{E}^t
=
-
\sum_{k=1}^N
\tilde{\mu}_k^t
\log(\tilde{\mu}_k^t+\epsilon),
\label{eq:spectral_entropy}
\end{equation}
and the effective diversity as
\begin{equation}
\mathcal{D}_{\mathrm{eff}}^t
=
\exp(\mathcal{E}^t).
\label{eq:effective_diversity}
\end{equation}
The normalized diversity is
\begin{equation}
D_{\mathrm{norm}}^t
=
\frac{\mathcal{D}_{\mathrm{eff}}^t}{N}.
\label{eq:normalized_diversity}
\end{equation}
When agents collapse into highly similar behaviors and similar Prospect States, $H_{\mathrm{fuse}}^t$ approaches a low-rank matrix and $\mathcal{D}_{\mathrm{eff}}^t$ becomes correspondingly small. When agents remain behaviorally and psychologically diverse, the effective diversity remains high. These metrics are computed periodically and used to assess heterogeneity preservation over long-horizon simulation.

\paragraph{Long-Horizon Simulation Loop.}
\label{subsec:multistep}
The simulation proceeds over a finite horizon $T$. The full simulation trajectory is
\begin{equation}
\tau = \bigl( \mathbf{M}^0, \mathcal{S}^0, \mathbf{a}^0, \mathbf{M}^1, \mathcal{S}^1, \ldots, \mathbf{M}^T, \mathcal{S}^T \bigr). \label{eq:trajectory}
\end{equation}

Algorithm~\ref{alg:psp} summarizes the entire simulation loop. At each time step, for each agent we optionally update the Semantic State every $K$ steps, then call the LLM to generate actions, execute market clearing, and update Prospect States with the Prospect-State Propagator. Heterogeneity is quantified by the spectral entropy $\mathcal{E}^t$ and effective diversity $\mathcal{D}_{\mathrm{eff}}^t$ of the fused Heterogeneity Matrix $H_{\mathrm{fuse}}^t$, which combines the Trajectory Matrix and the Prospect Matrix.

\begin{algorithm}[t]
\caption{Prospect-State Propagation Simulation Loop}
\label{alg:psp}
\begin{algorithmic}[1]
\STATE \textbf{Input:} $N$ agents, horizon $T$, update interval $K$, initial macro state $\mathcal{S}^0$
\STATE Initialize micro states $\{\mathcal{M}_i^0\}_{i=1}^N$ with profiles (basic, economic, psychological)

\FOR{$t = 0$ \TO $T-1$}

    \FOR{each agent $i$ \textbf{in parallel}}

        \IF{$t \bmod K = 0$}
            \STATE Update Semantic State 
            $\sigma_i^t \leftarrow 
            \mathrm{LLM}(s_i^t,\; e_i^t,\; 
            \mathrm{history}_{t-K+1:t},\; 
            \sigma_i^{t-K})$
        \ENDIF

        \STATE Generate economic action
        $a_i^t \sim 
        \pi_\theta(\cdot \mid 
        s_i^t,\; \sigma_i^t,\; 
        e_i^t,\; \mathcal{S}^t)$

    \ENDFOR

    \STATE Execute market clearing $\Psi$: 
    compute $\mathcal{S}^{t+1}$ and 
    record outcomes $\{o_i^t\}_{i=1}^N$

    \FOR{each agent $i$ \textbf{in parallel}}

        \STATE Update Prospect State
        $
        s_i^{t+1}
        =
        \Pi_{\mathcal{P}}
        (\Phi(s_i^t,\; o_i^t,\; \mathcal{S}^{t+1}))
        $

    \ENDFOR

    \STATE
    \raisebox{-0.12em}{
    \includegraphics[height=0.9em]{figures/yellow.pdf}}
    Construct $H_{\mathrm{traj}}^t$,
    \raisebox{-0.12em}{
    \includegraphics[height=0.9em]{figures/blue.pdf}}
    $H_{\mathrm{pros}}^t$,
    and
    \raisebox{-0.12em}{
    \includegraphics[height=0.9em]{figures/green.pdf}}
    $H_{\mathrm{fuse}}^t$;
    compute
    $\mathcal{E}^t$,
    $\mathcal{D}_{\mathrm{eff}}^t$,
    and
    $D_{\mathrm{norm}}^t$

\ENDFOR

\end{algorithmic}
\end{algorithm}

\subsection{Implementation Details}
\label{subsec:implementation}

Our implementation uses Qwen3-family models~\cite{qwen3}. The temperature is set to $0.7$ for decision calls and $0.2$ for summarization. The update interval $K$ is set to $5$. The Prospect-State Propagator coefficients are $\alpha_r=0.1$, $\beta_r=0.05$, $\eta_\lambda=0.001$, $\eta_\gamma=0.005$, $\zeta=0.1$, $\sigma_\lambda=0.02$. The threshold $E_{\text{thr}}=10$. Unless otherwise specified, each Prospect-State update is projected onto the predefined valid domain $\mathcal{P}$ after Eq.~\ref{eq:psp_rules}. For heterogeneity evaluation, we set the trajectory window to $L=12$ and use $\alpha=0.5$ in the fused Heterogeneity Matrix unless otherwise specified.

\section{Experiments}
\label{sec:experiments}

Our experiments answer the following research questions:

\begin{enumerate}[label=\textbf{RQ\arabic*:}, leftmargin=0pt, itemindent=0pt, labelsep=0.5em, align=left, wide=0pt]
    \item Why does increasing the number of agents fail to produce effective scaling in a state-of-the-art LLM-based economic simulation system?
    \item Does the proposed Prospect-State Propagation preserve agent heterogeneity at scale?
\end{enumerate}

\subsection{Experimental Setting}
\label{subsec:experimental_setting}

 \textit{Simulation Environment}. We simulate a closed economy where each agent makes labor supply and consumption decisions at each step. The macro state includes GDP, inflation, and unemployment.

\textit{Baseline Methods}. We compare four representative multi-agent architectures:

\begin{itemize}[leftmargin=1.5em]
    \item SaMAS: A situation-aware LLM-driven generative system for economic simulation, used as the strong prior system in our direct comparison \cite{chen2026empowering}.
    \item \mbox{\textsc{Summary System}}: Agents use plain summary: their histories are periodically compressed into LLM-generated summaries ($K=5$). The summary replaces raw event history in the context window \cite{park2023generative,qian2024chatdev}.
    \item \mbox{\textsc{Reflection System}}: Agents use periodic reflection: they maintain a memory stream and generate high-level reflections (``insights'') at regular intervals ($K=5$). Reflections are stored alongside recent events \cite{shinn2023reflexion,wang2024voyager}.
    \item \ours: Agents maintain a compact Prospect-State Vector $s_i^t \in \mathbb{R}^3$ updated via deterministic propagator at every step; Semantic State refreshed every $K=5$ steps.
\end{itemize}

\subsection{RQ1: Scaling Analysis of SaMAS and PspMAS}
\label{subsec:rq1}

We conduct a scaling analysis of \ours and SaMAS~\cite{chen2026empowering} using Qwen3-32B~\cite{qwen3} under the same evaluation protocol, including the same simulation horizon and comparable token budgets. We report results at two agent scales, $N=100$ and $N=500$, using Volatility Realism (VR) and normalized fused diversity $D_{\mathrm{norm}}$.

Table~\ref{tab:comparison} exposes a clear failure of the expected scaling law in SaMAS. A fivefold increase in nominal population, from $N=100$ to $N=500$, improves VR by only $0.2$ points ($81.8\%$ to $82.0\%$), while $D_{\mathrm{norm}}$ drops by $9.9$ points ($43.2\%$ to $33.3\%$). Performance therefore does not improve consistently with scale: the marginal gain in realism nearly vanishes, while the effective population becomes substantially more homogeneous. This result shows that nominal agent count is not equivalent to effective system scale when repeated semantic compression drives behavioral convergence. In contrast, \ours increases VR from $82.9\%$ to $85.3\%$ while retaining substantially higher diversity ($61.9\%$ at $N=100$ and $56.9\%$ at $N=500$). At $N=500$, it exceeds SaMAS by $3.3$ points in VR and $23.6$ points in $D_{\mathrm{norm}}$. This contrast indicates that preserving state heterogeneity is a prerequisite for realizing the expected benefits of scaling in MAS.

\begin{table*}[t]
\begin{minipage}[t]{0.48\textwidth}
\vspace*{7pt}
\centering
\setlength{\tabcolsep}{3pt}
\caption{Comparisons with the strong baseline SaMAS~\cite{chen2026empowering}.}
\label{tab:comparison}
\vspace{-8pt}
\begin{tabular}{c l c c}
\toprule
$N$ & System & VR (\%) $\uparrow$ & $D_{\mathrm{norm}}$ (\%) \\
\midrule
100 & SaMAS & $81.8$ & $43.2$ \\
100 & \ours & $\mathbf{82.9}$ & $\mathbf{61.9}$ \\
\midrule
500 & SaMAS & $82.0$ & $33.3$ \\
500 & \ours & $\mathbf{85.3}$ & $\mathbf{56.9}$ \\
\bottomrule
\end{tabular}
\end{minipage}\hfill
\begin{minipage}[t]{0.48\textwidth}
\vspace*{6pt}
\centering
\setlength{\tabcolsep}{3pt}
\caption{Normalized fused diversity $D_{\mathrm{norm}}$ (\%) under varying agent count $N$.}
\label{tab:agent_scale}
\begin{tabular}{c c c c}
\toprule
$N$ & \textsc{Summary} & \textsc{Reflection} & \ours \\
\midrule
20 & $68.2$ & $71.3$ & $73.8$ \\
50 & $40.3$ & $55.6$ & $67.5$ \\
200 & $21.2$ & $33.5$ & $59.6$ \\
500 & $20.1$ & $30.5$ & $56.5$ \\
\bottomrule
\end{tabular}
\end{minipage}

\par\vspace{1em}\noindent
\begin{minipage}[t]{0.48\textwidth}
\centering
\setlength{\tabcolsep}{4pt}
\caption{Normalized fused diversity $D_{\mathrm{norm}}$ (\%) under different LLM sizes.}
\label{tab:model_scale}
\begin{tabular}{l c c}
\toprule
Model & \textsc{Reflection} & \ours \\
\midrule
Qwen 3 - 8B & $32.1$ & $58.2$ \\
Qwen 3 - 32B & $36.5$ & $62.5$ \\
\bottomrule
\end{tabular}
\end{minipage}\hfill
\begin{minipage}[t]{0.48\textwidth}
\centering
\setlength{\tabcolsep}{3pt}
\caption{Normalized fused diversity $D_{\mathrm{norm}}$ (\%) under different simulation horizons.}
\label{tab:horizon}
\begin{tabular}{c c c c}
\toprule
Horizon & \textsc{Summary} & \textsc{Reflection} & \ours \\
\midrule
50 & $28.5$ & $42.5$ & $61.2$ \\
200 & $18.5$ & $30.2$ & $57.8$ \\
\bottomrule
\end{tabular}
\end{minipage}
\end{table*}

\subsection{RQ2: Heterogeneity Preservation at Scale}
\label{subsec:rq2}

To quantify whether \ours preserves agent heterogeneity as the system scales, we compute the normalized diversity $D_{\mathrm{norm}} = \mathcal{D}_{\mathrm{eff}} / N$ from the fused Heterogeneity Matrix $H_{\mathrm{fuse}}^t$ defined in Section~\ref{subsec:scaling}. This matrix combines the Trajectory Matrix, which measures recent action diversity, and the Prospect Matrix, which measures path-dependent psychological diversity. Higher $D_{\mathrm{norm}}$ indicates more behaviorally distinct agents.

\paragraph{Varying agent count.}
Table~\ref{tab:agent_scale} shows $D_{\mathrm{norm}}$ as a function of $N$. \mbox{\textsc{Summary System}} saturates at $N \approx 200$, confirming that repeated semantic compression homogenizes both recent action trajectories and prospect-state traces regardless of population size. \mbox{\textsc{Reflection System}} shows sublinear scaling. In contrast, \ours maintains high fused diversity even at $N=500$, demonstrating stronger heterogeneity scaling.

\paragraph{Varying model size.}
Table~\ref{tab:model_scale} examines the effect of LLM backbone size on heterogeneity preservation. Scaling from Qwen3-8B to Qwen3-32B increases $D_{\mathrm{norm}}$ from $32.1\%$ to $36.5\%$ for \mbox{\textsc{Reflection System}} and from $58.2\%$ to $62.5\%$ for \ours.

\paragraph{Varying simulation horizon.}
Table~\ref{tab:horizon} summarizes $D_{\mathrm{norm}}$ at simulation horizons 50 and 200. \mbox{\textsc{Summary System}} degrades rapidly as repeated compression accumulates information loss. \mbox{\textsc{Reflection System}} shows more graceful degradation but still suffers from cumulative drift. \ours maintains remarkably stable fused diversity, indicating that the Prospect-State Propagator preserves both behavioral trajectories and path-dependent prospect-state differences over long horizons.

\subsection{Summary of Findings}
\label{subsec:summary}

\begin{enumerate}[label=\textbf{RQ\arabic*:}, leftmargin=0pt, itemindent=0pt, labelsep=0.5em, align=left, wide=0pt]
    \item Failure of scaling in SOTA MAS: Increasing the SaMAS population fivefold, from $N=100$ to $N=500$, improves VR by only $0.2$ points while reducing $D_{\mathrm{norm}}$ from $43.2\%$ to $33.3\%$. In contrast, \ours improves VR from $82.9\%$ to $85.3\%$ while retaining much higher diversity ($61.9\%$ to $56.9\%$), showing that heterogeneity enables effective scaling.
    
    \item Heterogeneity preservation at scale: At $N=500$, \ours maintains $D_{\mathrm{norm}} > 0.56$, whereas \mbox{\textsc{Summary System}} saturates near $N=200$ ($D_{\mathrm{norm}} \approx 0.21$). This advantage persists across LLM sizes and simulation horizons.
\end{enumerate}

\section{Related Work}
\label{sec:relatedwork}

\noindent\textbf{Macroeconomic Simulation}
Traditional macroeconomic models such as DSGE \cite{smets2007shocks,clarida1999science,christiano2005} and VAR \cite{sims1980macroeconomics,del2015inflation} primarily analyze economic fluctuations and policy transmission in a top-down manner through mathematical modeling of the relationship between ``representative agents'' and macroeconomic variables. Agent-based Modeling adopts a ``bottom-up'' paradigm for social simulation: by simulating the behaviors and interactions of micro-level individuals within a specific environment, it reproduces macroeconomic phenomena, thereby overcoming the limitation of the ``representative agent'' assumption inherent in traditional models \cite{acemoglu2012network,poledna2023economic,caiani2016agent,axtell2001zipf,geanakoplos2012getting,dawid2018agent}. In recent years, LLM-driven agents have endowed ABM with scenario-based interaction capabilities, allowing for the simulation of more complex economic behaviors and enabling more sophisticated economic simulations \cite{yang2026twinmarket,li2024econagent,chen2024agentverse,li2024mars,hagendorff2025deception,argyle2025llm,jia2024can,widler2026investigation}. Despite these advances, existing approaches largely rely on periodic semantic summarization or reflection to compress long economic histories, a practice that risks attenuating fine-grained behavioral trajectories. This compression, though token-efficient, tends to homogenize agents over time by progressively smoothing away their early fluctuations.

\noindent\textbf{LLM-driven Agentic System}
 AI Agent is an autonomous system that perceives its environment and takes actions to achieve goals. These actions are not hard-coded but rather emerge from situated interactions, continuously evolving through feedback from interactions with the environment and other agents. This paradigm has achieved notable success in multi-turn conversation systems \cite{zheng2023judging,wu2024autogen}, code generation \cite{jimenez2024swe,yang2024swe}, and general-purpose assistant tasks \cite{wang2024rolellm,chen2024agentverse,chen2025diffvsgg,chen2024gvseg,chen2024pipa++}. Furthermore, the same paradigm has advanced social reasoning \cite{gandhi2023understanding,sap2019social,mankowitz2023faster,zhang2024llm}, policy optimization \cite{silver2017mastering,silver2016mastering}, and behavioral simulation \cite{chen2024persona,park2023generative,jia2024can,li2024agent}. This paradigm excels at leveraging LLMs for flexible, context‑aware reasoning, enabling agents to adapt to novel situations without hand‑crafted rules. The feedback‑driven, interactive design also allows agents to continuously refine their strategies, producing emergent behaviors that static models cannot replicate. As individual agent capabilities improve, LLM-driven Multi-agent Systems have come into sharper focus, and recent research has begun to further investigate the collective behavior and emergent dynamics arising from multi-agent interactions. Recent work demonstrates that LLM-based multi-agent systems can spontaneously form social norms, collaborative structures, and complex group behaviors through ongoing interaction \cite{li2025language,schneider2025learning,wu2024shall,riedl2025emergent}. Moreover, recent efforts have leveraged these systems for macroeconomic forecasting \cite{jin2024time,ansari2024chronos,rasul2023lag,garza2023timegpt}, revealing intricate emergent macro-level phenomena.

\section{Conclusion}
\label{sec:conclusion}
We present \ours, the first macroeconomic simulation MAS that incorporates prospect theory into economic state updating. To achieve this, we first decouple each agent’s micro state into two complementary components. We then leverage the Prospect State to continuously inject heterogeneity into the system, while using the Semantic State to generate human-like actions. By combining the two, \ours not only significantly reduces LLM inference token consumption, making effective scaling possible, but also achieves strong simulation performance. More broadly, our work provides new insights into addressing the challenges of LLM-driven MAS through classic theories from behavioral economics. Future work could extend \ours to other complex socio-economic domains such as financial market regulation, climate policy negotiation, and organizational behavior simulation, where bounded rationality and heterogeneous decision-making play critical roles.

\section*{Limitations}
This study has several limitations. First, the current simulations are still conducted within a simplified closed economy environment, and thus may fail to capture open economy effects, institutional constraints, or real-world policy frictions. Second, decisions based on large language models may inherit biases from their underlying models; future research should examine robustness across different model families, prompting strategies, and calibration settings. Third, during the training process, large language models absorb cultural, political, and ideological biases from the internet, academic literature, and policy documents. It is currently impossible to completely disentangle these deep-seated value orientations, posing a serious challenge for applications seeking "value-neutral" policy insights in practice.

\bibliography{custom}

\end{document}